\documentstyle[12pt,epsfig]{article}

\begin{document}

 \setlength{\baselineskip}{16.0pt}

\begin{center}
{\bf Bohmian
trajectories on a toroidal surface} \\ {\vskip 6pt} Mario Encinosa \footnote{ corresponding author}  and Fernando Sales-Mayor \\
Department of Physics \\  Florida A \& M University \\
  Tallahassee, Florida 32307
  \\encinosa@cennas.nhmfl.gov, fsm@lanczos.cm.utexas.edu
\end{center}
 \setlength{\baselineskip}{16.0pt}
\begin{abstract} \noindent Bohmian trajectories
on the toroidal surface $T^2$ are determined from eigenfunctions of the
Schrodinger equation. An expression for the monodromy matrix $M(t)$ on a
curved surface is developed and eigenvalues of $M(t)$ on $T^2$
calculated. Lyapunov exponents  for trajectories on $T^2$  are found for
some trajectories to be of order unity.


\end{abstract}
\vskip 6pt

 The  equation of motion for calculating Bohmian trajectories  follows from inserting
 $\Psi({\bf r},t)=R({\bf r},t)e^{iS({\bf r},t)}$
into the Schrodinger equation and separating the ensuing expression into
real and imaginary parts [1]. The result is ($m=\hbar = 1$)
$$
 {d {\bf v} \over {dt}} = - \nabla (V + Q) \eqno(1)
$$
with ${\bf v} = \nabla S$ and $Q$ the quantum potential
$$
Q= -{1 \over 2}{ \nabla^2 R \over R}. \eqno(2)
$$
 Given a distribution of initial
positions as determined from $\psi^* \psi$ the standard results of
quantum mechanics are recovered.

 This letter  concerns the influence of curvature on Bohmian
trajectories. Wu and Sprung [2] have pointed to the importance of
 a trajectory's initial position as critical to its
 evolution in time. If the trajectory lies on a curved surface,
 local curvature near the trajectory's initial position
 may be expected to play a role in its character. Here wave
functions on a curved surface are employed to generate trajectories
 from $S$ [2,3,4].

 $T^2$ was chosen  as the surface with
 which to investigate curvature
effects on Bohmian trajectories
 for several reasons: First, the torus has
 non-trivial mean and Gaussian curvatures[5].
 Second, good approximate wave functions for a particle on
$T^2$ are available [6].
 Finally,  the rectangular strip
 $R^1 {\rm x} R^1$ with periodic boundary conditions can be used as a flat torus
analog for  comparison  between trajectories on it and those on $T^2$.
For
 convenience the strip  will be referred to as $F^2$.

A toroidal surface with major radius $R$ and minor radius $a$ may be
characterized by the Monge form
$$
{\bf r}(\theta,\phi)=(R+a \ {\rm cos} \theta){\bf e}_{\rho}+a\ {\rm sin}
\theta\ {\bf e}_{z}\eqno(3)
$$
for which
$$
ds^2=a^2  d\theta^2 + (R + a \ {\rm \cos} \theta)^2  d\phi^2.
 \eqno(4)
$$
The Hamiltonian is taken as $H = -{1 \over 2}\nabla^2$.  Defining $\alpha
= {a \over R},$ $\ $ $\beta = 2Ea^2$, and making the standard $\chi(\phi)
= e^{im\phi}$  ansatz for the azimuthal eigenfunction gives the
Schrodinger equation
$$
  {\partial^2 \psi \over \partial \theta^2}-
 {\alpha \ {\rm sin}\ \theta \over [1 + \alpha \ {\rm \cos}\  \theta]}{\partial \psi \over \partial \theta}
-{m^2 \alpha^2 \over [1 + \alpha \  {\rm \cos}\theta]^2}\psi +\beta\psi =
0. \eqno(5)
$$
Eq. (5) was solved in [6] by a Fourier method. The explicit forms of the
surface toroidal wave functions (STWs) used here are given in table I for
$R=1, a= 1/2$. The eigenfunctions on $F^2$ analogous to those in table I
are $\bigg ( {{{\rm cos}n\theta} \atop {{\rm sin} n\theta}}\bigg
)e^{im\phi}$.

As evidenced in table I, six  STW  states were chosen to comprise three
positive parity  states and three negative parity  states. The $[nm]$
values selected for the negative parity states are the same as those
selected for the positive parity states.  This $[nm]$ matching
 was motivated by a desire to construct (as closely as possible) functions
 for comparison to
$ e^{in\theta}$  functions on $F^2$. Additionally, without superpositions
of the form $\psi_{nm}^+ \pm i\psi_{nm}^-$ some interesting motion about
the minor radius of $T^2$ would not be manifest.

The phase $S$ for $T^2$ and $F^2$ is generated with a six state
superposition
$$
\Psi(\theta ,\phi,t)=\sum_{nm}c_{nm}\Psi_{nm}(\theta,\phi)e^{-iE_{mn}t}.
\eqno(6)
$$
Trajectories were determined from
$$
S = tan^{-1} \bigg [ {{{Im \Psi} \over {Re \Psi} }\bigg] }
 \eqno (7)
$$
 and
$$
{d {\bf r}\over dt}  = \nabla S. \eqno(8)
$$

Eqs. (7) and (8) yield many classes of surface trajectories. Figures 1
and 2 give two state  results for $T^2$ to the left of each figure and
for $F^2$ to the right. Figure 3 shows a quantized trajectory structure
that emerges from a combination of negative parity states on $T^2$.
Figure 4 is a $(\theta,\dot{\theta})$
 plot for a path on $T^2$ showing rapid variation in phase space. Figure
5 is an example of a $(\theta,\dot{\theta})$ plot for a superposition
with a dominant mode and  small admixtures of two other states. A small
change on the order of a few parts in $10^{-3}$ in $\theta_0$ causes
ample modification to the path. $F^2$ phase space plots have not been
shown because they demonstrate (at least for the cases above) very little
structure.

A measure of the divergence of two initially nearby trajectories is
associated with the eigenvalues of the monodromy matrix $M(t)$ [7,8],
used to good effect by Frisk [3] for the study of Bohmian paths on a
two-dimensional rectangular geometry. $M(t)$ gives the time evolution of
the separation vector $\delta {\bf x}$ via
$$
\delta {\bf x}(t)=M(t)\delta {\bf x}(0). \eqno(9)
$$
Since
$$
{d \over {dt}}[\delta {\bf x}(t)]=\nabla [\delta S], \eqno(10)
$$
the monodromy matrix can be shown to obey the differential equation
$$
{dM \over {dt}}=JM. \eqno(11)
$$
For $R^1 {\rm x} R^1$ with or without periodic boundary conditions the
$J$ matrix is [3]
$$
J = \pmatrix{ S_{xx} & S_{xy} \cr S_{yx} & S_{yy} \cr} . \eqno(12)
$$
On a two dimensional surface characterized by a locally orthogonal metric
$$
ds^2=g_{uu}du^2+g_{vv}dv^2
$$
Eq. (12) must be extended to (written in a cumbersome but illustrative
form)
$$
J = \pmatrix{ {\sqrt g^{uu}} S_{uu} {\sqrt g^{uu}} & {\sqrt g^{uu}}
S_{uv}{\sqrt g^{vv}} \cr {\sqrt g^{vv}} S_{vu} {\sqrt g^{uu}} & {\sqrt
g^{vv}} S_{vv} {\sqrt g^{vv}} \cr} . \eqno(13)
$$
On $T^2$ Eq. (13) yields
$$
J = \pmatrix {{S_{\theta\theta}}\over{\alpha^2} &
{S_{\theta\phi}}\over{\alpha G} \cr {S_{\phi\theta}}\over{\alpha G} &
{S_{\phi\phi}}\over{G^2} \cr}  \eqno(14)
$$
with $G \equiv 1+\alpha \ {\rm cos}\ \theta$.

It is worth noting that for curved surfaces $\delta$ and $\nabla$ do not
commute. The ordering chosen in Eq. (10) insures all quantities lie on
the local tangent space [9,10].

The differential equations implicit in Eq. (11) were found to be  time
consuming and slow to converge to
 suitable accuracy with standard methods on $T^2$.  Further, it
 was not clear that the results were accurate to any order.
 However, the foremost  goal here
 was  obtaining comparisons between the eigenvalues of $M(t)$ on
 $T^2$ versus those on $F^2$ for several values of ${\theta_0}$.
With this is mind, a Mathematica code was set to solve the differential
equations to a lower
 accuracy for the relatively short time $t=10$. The advantages of adopting this procedure
 were a) confidence that our numbers were accurate to at least four
 significant digits
  and b) each point took  only at most two minutes to
  acquire on an 800 MHz rated home PC [11].

 Rather than
work directly with the eigenvalues $\beta_i$ of $M(t)$,  a Lyapunov
exponent $\lambda_i$ is defined through the relation
$$
\lambda_i(t)= {1\over t}\ln \beta_i(t). \eqno(15)
$$
 A more
convenient definition which serves to dispense of an unknown overall norm
is one  employed in quantum Monte Carlo calculations [12] and  adopted
here,
$$
\lambda_i = {1\over {(t_2-t_1)}}\ln {{\beta_i(t_2)} \over \beta_i(t_1)}.
\eqno(16)
$$
The larger of the two  values of $\lambda_i \equiv \lambda$ was taken.

Tables II and III give values of $\lambda$ as a function of $\theta_0$
for two sets of $c_{nm}$ values. For both sets $\lambda$ is generally an
order of magnitude larger on $T^2$ than on $F^2$ (save for one anomalous
point). The $F^2$ values show a symmetry about $\theta = \pi$ while the
$T^2$ values do not. The $T^2$ results certainly show dependence on
$\theta_0$ but do not show any simple relation to the Gaussian or mean
curvatures $K$ and $H$ on $T^2$. This is not unexpected. The Hamiltonian
of Eq.(5)
 does not incorporate $H$ and $K$ explicitly but rather factors
related to them in a complicated manner. A trajectory on $T^2$ generated
from a representative point in table II is given in figure 6, and one
from table III is shown in figure 7. Again, the corresponding $F^2$ plots
do not possess enough structure to warrant their inclusion.

In this letter linear combinations of  wave functions were employed to
generate surface and phase space plots of Bohmian trajectories on $T^2$
and $F^2$. The plots and tables illustrate that curvature  can  alter
trajectory structure. This is a direct manifestation of the form of the
wave functions on a curved surface. Each exponential function in a
superposition of states on $F^2$ possesses its own time dependent energy
phase while each STW is a superposition of several functions attached to
one phase.

$c_{nm}$ combinations making  Lyapunov exponents on $T^2$
  appreciably smaller than those on $F^2$ over many values of $\theta_0$
  have not been found.
  Positive $\lambda$ of order unity
is usually taken to indicate  chaotic behavior, but while some
trajectories are certainly complex, and $\lambda$ an adequate measure of
the distortion of local tangents, we consider it premature to state the
trajectories are chaotic [13,14]. Nevertheless, it has been shown that a
few low-lying states  can yield complex
 phase space behavior on the the torus. Because the torus is a  simple
 compact surface, its curvature  likely causes less modification to trajectory
 structure than surfaces with rapidly varying regions of
$H$ and $K$. Those surfaces are certain to induce  greater
 complexity in the trajectories.

\vfill \eject \centerline{\bf Acknowledgments} \vskip 6pt
 The authors would like to thank Todd Timberlake for useful
discussions. M.E. was partially supported by the NASA grant NAG2-1439 and
F.S-M. by the Army High Performance Computing Research Center grant,
cooperative agreement number DAAD 19-01-2-0014. \vskip 12pt

\centerline{\bf References} \vskip 6pt \noindent 1. P. R. Holland, ${\it
The \ quantum \ theory \ of \ motion}$ (Cambridge University Press,
Cambridge, 1993). \vskip 6pt \noindent 2. J. Wu and D.W.L. Sprung, Phys.
Lett. A ${\bf 261}$, 150 (1999). \vskip 6pt \noindent 3. H. Frisk, Phys.
Lett. A ${\bf 227}$, 139 (1997). \vskip 6pt \noindent 4. R.H. Parmenter
and R.W. Valentine, Phys. Lett. A ${\bf 201}$, 1 (1995). \vskip 6pt
\noindent 5. http://mathworld.wolfram.com/Torus.html \vskip 6pt \noindent
 6. M. Encinosa and B. Etemadi, quant-ph 0200501 and submitted to Found. Phys. Lett.
  \vskip 6pt \noindent 7. F.H.M. Faisal and U. Schwengelbeck, Phys.
Lett. A ${\bf 207}$, 31 (1995).  \vskip 6pt \noindent 8.  U.
Schwengelbeck and F.H.M. Faisal, Phys. Lett. A ${\bf 199}$, 281 (1995).
\vskip 6pt \noindent 9. R.W.R. Darling, ${\it Differential \ forms \ and
\ connections}$ (Cambridge University Press, Cambridge, 1994). \vskip 6pt
\noindent 10. Reversing the order of $\nabla$ and $\delta$ and projecting
onto the tangent plane does not lead to a linear relation between ${\bf
\delta} x(t)$ and ${\bf \delta} x(0)$.
 \vskip 6pt
\noindent 11. The  authors will supply the Mathematica source code upon
request.  \vskip 6pt \noindent  12. D. Ceperley and M.H. Kalos, ${\it
Monte \ Carlo\ methods \ in \ statistical \ physics}$, K. Binder, Ed.
(Springer Verlag Berlin, 1979). \vskip 6pt \noindent 13. D. D\"{u}rr, S.
Goldstein and N. Zanghi, J. Stat. Phys. ${\bf 68}$, 259 (1992). \vskip
6pt \noindent 14. M.C. Gutzwiller, ${\it Chaos \ in \ classical \ and \
quantum \ mechanics}$ (Springer-Verlag New York Inc. 1990).
 \vfill \eject

\begin{center}
{\bf Figure captions}
\end{center}

\vskip 12pt \noindent Fig. 1. Trajectories for the superposition
$\Psi(\theta,\phi,t)=\sqrt {2\over  3}\Psi_{32}^+
 + i  \sqrt{1\over  3}\Psi_{32}^- $ with  $t=30$ and $\theta_0 = 0$.
  The $T^2$ trajectory
appears to the left of the figure. The $F^2$ trajectory projected onto
the torus is to the right.

 \vskip 12pt \noindent Fig. 2.
Trajectories for  $\Psi(\theta,\phi,t)= \sqrt{2\over  3}\Psi_{32}^-
 + i \sqrt {1\over  3}\Psi_{21}^+ $ with  $t=24$ and $\theta_0 = 1.05\pi$.

\vskip 12pt \noindent Fig. 3. Trajectories for $\Psi(\theta,\phi,t)=
\sqrt{5.2\over  12}\Psi_{21}^+ +(\sqrt {1.4\over  12}\Psi_{21}^- -
i\sqrt{5.2\over  12})\Psi_{21}^-$ with  $t=50$ and $\theta_0 = 0$.

\vskip 12pt \noindent Fig. 4. $(\theta,\dot{\theta})$ plot for
$\Psi(\theta,\phi,t)= \sqrt{1\over  3}\Psi_{21}^+ +i\sqrt {1\over
3}\Psi_{32}^+ - i\sqrt{1\over 3}\Psi_{32}^-$ with  $t=30$ and $\theta_0 =
1.25\pi$.

\vskip 12pt \noindent Fig. 5. $(\theta,\dot{\theta})$ plots for
$\Psi(\theta,\phi,t)= \sqrt{0.02\over  12}\Psi_{10}^- +\sqrt {11.96\over
12}\Psi_{21}^- +\sqrt{.02\over  12}\Psi_{32}^-$ with  $t=30$ and
$\theta_0 = 1.424\pi$ shown on the left and for $\theta_0 = 1.429\pi$
shown on the right.

\vskip 12pt \noindent Fig. 6. $(\theta,\dot{\theta})$ plot for
$\Psi(\theta,\phi,t)= \sqrt{1\over  2}\Psi_{32}^+ +\sqrt {1\over
2}\Psi_{32}^-$ with  $t=36$ and $\theta_0 = 0$.

\vskip 12pt \noindent Fig. 7. $(\theta,\dot{\theta})$ plot for
$\Psi(\theta,\phi,t)= \sqrt{1\over  2}\Psi_{10}^+ +i\sqrt {1\over
2}\Psi_{10}^-$ with  $t=36$ and $\theta_0 = 0$.

\vfill \eject

\begin{table}
\caption{Surface torodial wave functions and eigenvalues for $R=1,a=1/2$.
Coefficients not listed are at least an order of magnitude smaller than
those given. }
\begin{center}
\begin{tabular}{|l|l|}
\hline
\ \qquad  \qquad \qquad  \qquad $\Psi^{\pm}_{nm};  R=1,a=1/2$ & $\ \ \ \beta$\\
\hline
$\Psi^{+}_{10}=-.2176+.4352 \ \rm cos \theta - .0714\ \rm cos 2\theta +.0118 \ \rm cos 3\theta$ &1.2223 \\
 $\Psi^{+}_{21}=-.0733+.2419 \ \rm cos \theta-.8393\ \rm cos2 \theta+.0541\ \rm cos3 \theta$ &4.4767 \\
 $\Psi^{+}_{32}=-.0420+.1240 \ \rm cos \theta-.2772\ \rm cos 2\theta+.8240 \ \rm cos 3\theta$ &10.6657 \\
 $\Psi^{-}_{10}=+.8118 \ \rm sin\theta-.0739\ \rm sin2\theta$ & 0.9767\\
 $\Psi^{-}_{21}=-.1799 \ \rm sin\theta-.8367\ \rm sin2\theta$ & 4.4106  \\
 $\Psi^{-}_{32}=-.0808\ \rm sin\theta+.2568\ \rm sin2\theta-.8257\ \rm sin3\theta$ & 10.6151 \\
 \hline
\end{tabular}
\end{center}
\end{table}

\begin{table}
\caption{Lyapunov exponents $\lambda_9$ measured at $t=9$, $\lambda_{10}$
measured at $t=10$ and $\lambda$ as defined by Eq. (16) for the
superposition $\sqrt {1\over 2}\Psi^{+}_{32}+\sqrt {1\over
2}\Psi^{-}_{32}$. The upper half of the table gives results for $T^2$ and
the lower half are results for $F^2$.}
\begin{center}
\begin{tabular}{|c|cccccccccccc|}

\hline $\theta_0$ & 0 & $\pi \over 6$ & $\pi \over 3$ & $\pi \over 2$ &
$2 \pi \over 3$ & $5 \pi \over 6$ & $\pi$ & $ 7\pi \over 6$ & $4 \pi
\over 3$& $3 \pi \over 2$ &$5\pi \over 3$ &
$11 \pi \over 6$\\
\hline

$\lambda_9$ &2.12 & 4.23 &2.53 & 1.91 & 3.36 & 2.29 & 2.75 &2.55 &3.11 &1.72 &1.23 &2.70  \\
$\lambda_{10}$ &4.10 & 5.08 &3.67 & 2.65 & 4.17 & 2.69 & 2.57 &2.85 &5.73 &2.39 &2.59 &3.27 \\
$\lambda$  &21.9 & 12.7 &13.9 &9.35& 11.5& 6.35 & .96 & 5.53 &29.3 &8.44 &14.8 &8.40\\
\hline
$\lambda_9$ &.030 & .036 &.034 & .033 & .036 & .032 & .030 &.036 &.034 &.033 &.033 &.036  \\
$\lambda_{10}$ &.047 & .056 &.050 & .046 & .049 & .048 & .047 &.056 &.050 &.046 &.046 &.056\\
$\lambda$  &.185 & .234 &.179 & .161 & .155 & .195 & .186 & .233 &.179 &.160 &.155 &.235\\
 \hline
\end{tabular}
\end{center}
\end{table}

\begin{table}
\caption{Lyapunov exponents $\lambda_9$ measured at $t=9$, $\lambda_{10}$
measured at $t=10$ and $\lambda$ as defined by Eq. (16) for the
superposition $\sqrt {1\over 2}\Psi^{+}_{10}+\sqrt {1\over
2}\Psi^{-}_{10}$. The upper half of the table gives results for $T^2$ and
the lower half are results for $F^2$. }
\begin{center}
\begin{tabular}{|c|cccccccccccc|}

\hline
 $\theta_0$ & 0 & $\pi \over 6$ & $\pi \over 3$ &
$\pi \over 2$ & $2 \pi \over 3$ & $5 \pi \over 6$ & $\pi$ & $ 7\pi
\over 6$ & $4 \pi \over 3$& $3 \pi \over 2$ &$5\pi \over 3$ &
$11 \pi \over 6$\\
\hline
$\lambda_9$ &2.46 &1.69 &1.68 & 1.63 &1.67 & 1.86 & 2.01 &3.59 &3.64 &3.22 &3.13 &6.87  \\
$\lambda_{10}$ &2.55 & 1.65 &1.63 & 1.58& 1.61& 1.74 & 1.92 &3.28 &3.97 &3.99 &3.73 &6.90 \\
$\lambda$  &3.39 & 1.23 &1.02 &1.19& 1.07 & .067 & 1.18 & .447 &7.00 &10.90 &9.14 &7.16\\
\hline
$\lambda_9$ &.086 & .684 & -.003& .022 & .002 & .009 & .086 & .684 & -.003& .022 & .002 & .009 \\
$\lambda_{10}$ &.122 & .336 & .450  & .124 & .015 & .032 & .122 & .336 & .450  & .124 & .015 & .032\\
$\lambda$ &.415 & -2.45& 4.52 & .942 & .133 & .239 & .415  & -2.45& 4.52 & .942 & .133 & .239  \\
 \hline
\end{tabular}
\end{center}
\end{table}

\eject

\begin{figure}
\begin{center}
\leavevmode \epsffile{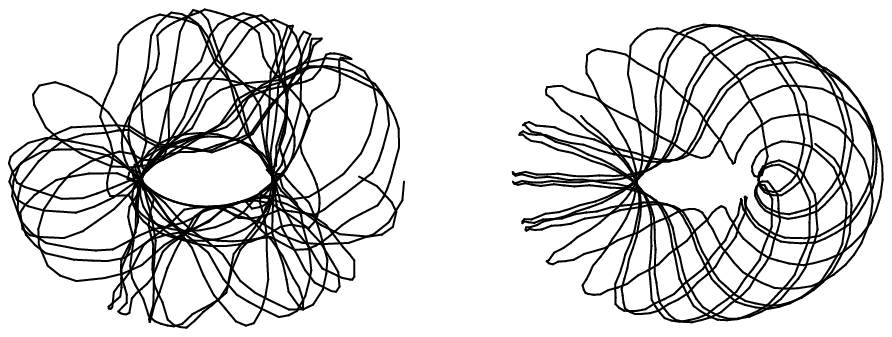}
\end{center}
\caption{}
\end{figure}

\begin{figure}
\begin{center}
\leavevmode \epsffile{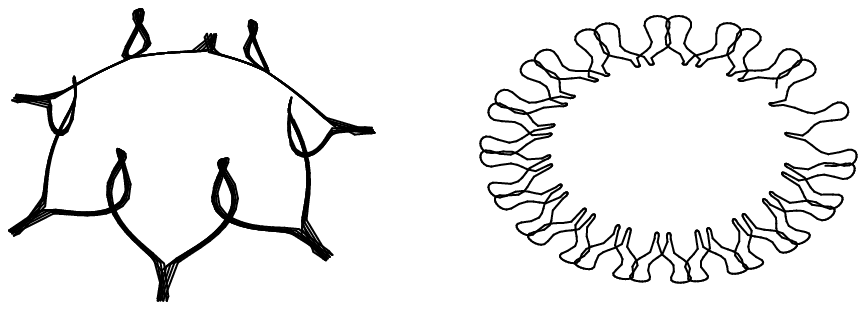}
\end{center}
\caption{}
\end{figure}

\begin{figure}
\begin{center}
\leavevmode \epsffile{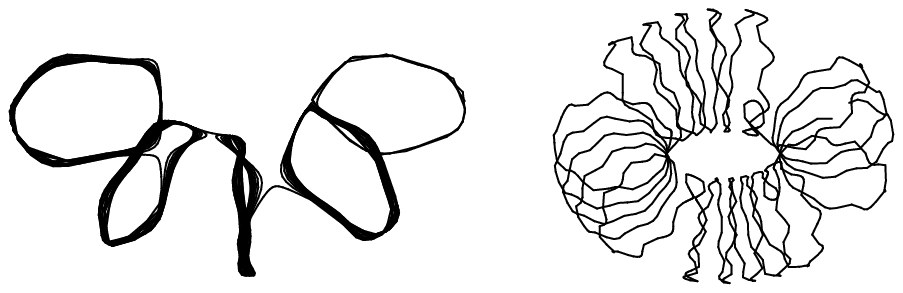}
\end{center}
\caption{}
\end{figure}

\begin{figure}
\begin{center}
\leavevmode \epsffile{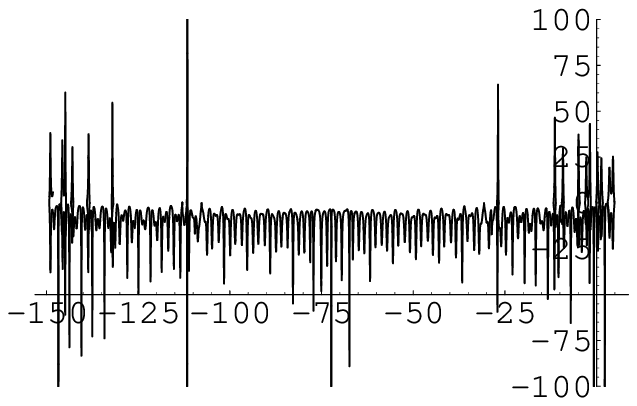}
\end{center}
\caption{}
\end{figure}

\begin{figure}
\begin{center}
\leavevmode \epsffile{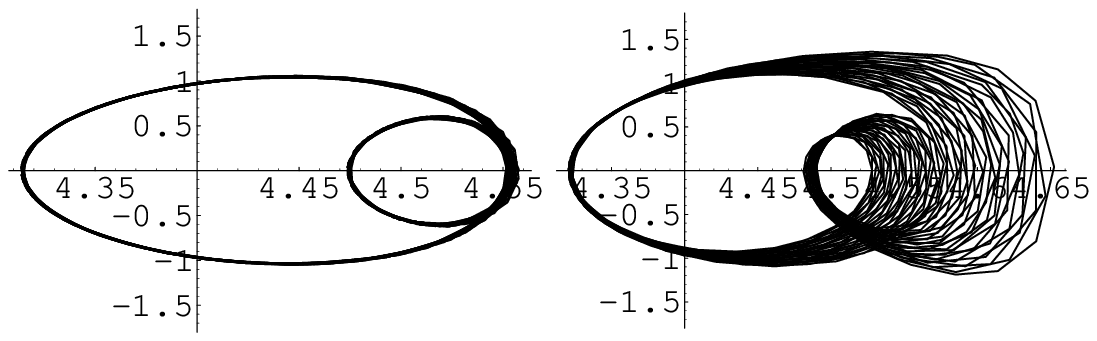}
\end{center}
\caption{}
\end{figure}

\begin{figure}
\begin{center}
\leavevmode \epsffile{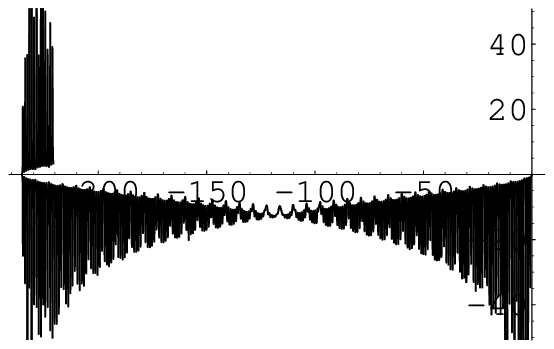}
\end{center}
\caption{}
\end{figure}

\begin{figure}
\begin{center}
\leavevmode \epsffile{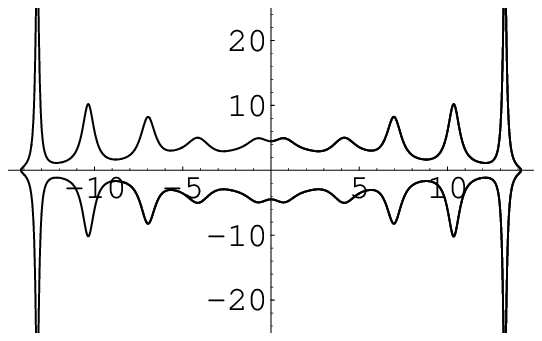}
\end{center}
\caption{}
\end{figure}

\end{document}